# MECHANISM OF WALL TURBULENCE IN BOUNDARY LAYER FLOW

HUA-SHU DOU

*Temasek Laboratories, National University of Singapore,
Singapore 117508
tsldh@nus.edu.sg; huashudou@yahoo.com*

BOO CHEONG KHOO

*Department of Mechanical Engineering, National University of Singapore,
Singapore 119260
mpekbc@nus.edu.sg*

The energy gradient method is used to analyze the turbulent generation in the transition boundary layer flow. It is found that the maximum of the energy gradient function occurs at the wall for the Blasius boundary layer flow. At this location under a sufficiently high Reynolds number, even a low level of free-stream disturbance can cause the turbulent transition and sustain the flow to be in a state of turbulence. This is an excellent explanation of the physics of self-sustenance of wall turbulence. The mechanism of receptivity for boundary layer flow can also be understood from the energy gradient criterion. That is, the free-stream disturbance can propagate towards the wall by the "energy gradient" process to cause turbulent transition, and the transition point in boundary layer can be moved forward towards the leading edge when the level of external disturbance increases.

## 1. Introduction

Flow stability and turbulent transition are challenging research topics for modern fluid dynamics. However, these problems are still poorly understood although more than a century effort has been made [1-3]. Since the earlier years of turbulence investigation, the origin of turbulence in the boundary layer has remained a key challenge: whether it originates from the wall or comes from the external freestream turbulence. Progress in recent years has shown that there is a self-sustenance mechanism of wall turbulence which is almost not affected by the outer flow [2], while this phenomenon does not occur for Poiseuille flows. However, it is also found that a "receptivity" mechanism exists for boundary layer flows[3], which characterizes the effect of freestream turbulence. So, what is the relation between these two phenomena, if any? In this study, the energy gradient method in [4-5] is used to analyze the turbulent generation in the transition in boundary layer flows. Based on these, the physics of self-sustenance of wall turbulence (which is absent in Poiseuille flow) and the effect of free-stream turbulence are discussed.

## 2. Energy Gradient Theory Applied to Boundary Layer Flow

To clarify the mechanism of flow instability and turbulent transition, Dou [4] has shown that the relative magnitude of the total mechanical energy of fluid particles gained and the energy loss due to viscous friction in a disturbance cycle determines the disturbance amplification or decay. For a given flow, a stability criterion is written as [4-5],

$$K \frac{v'_m}{U} < Const, \quad K = \frac{(\partial E / \partial n)}{(\partial H / \partial s)},  \tag{1}$$

where $K$ is a dimensionless field variable (function) and expresses the ratio of the gradient of the total mechanical energy in the transverse direction and the loss of the total mechanical energy in the streamwise direction. Here, $E = p + 0.5\rho V^2$ is the total mechanical energy per *unit volumetric fluid*, $s$ is along the streamwise direction, $n$ is along the transverse direction, $H$ is the energy loss per *unit volumetric fluid* along the streamline, $\rho$ is the fluid density, u is the streamwise velocity of main flow, U is the characteristic velocity, and $v'_m$ is the amplitude of disturbance velocity. Since the magnitude of $K$ is proportional to the global Reynolds number ($\mathrm{Re} = \rho U L / \mu$) for a given geometry [4-5], the criterion of Eq.(1) can be written as,

$$\mathrm{Re} \frac{v'_m}{U} < Const \quad \text{or} \quad (\frac{v'_m}{U})_c \sim (\mathrm{Re})^{-1}.  \tag{2}$$

This scaling has been confirmed by careful experiments observed for the pipe flow, and this result is in agreement with the asymptotic analysis of the Navier-Stokes equations (for $\mathrm{Re} \to \infty$) [4-5].

The boundary layer flow along a flat plate at zero incidence angle is shown in Fig.1. The leading edge of the plate is at x=0 and the plate is parallel to the x-axis and infinitely long downstream. The steady flow with a free-stream velocity, $u_\infty$, is parallel to the x-axis. The velocity of the external inviscid flow at the edge of the boundary layer is constant. Assuming that the principle of similarity is applicable to the boundary layer, a dimensionless coordinate and stream function can be introduced as [1],

$$\eta = y\sqrt{\frac{u_\infty}{vx}}, \quad \psi = \sqrt{vxu_\infty} f(\eta).  \tag{3}$$

The velocities are obtained as

$$u = u_\infty f'(\eta), \quad v = \frac{1}{2}\sqrt{\frac{vu_\infty}{x}}(\eta f' - f).  \tag{4}$$

Then, the boundary layer equations become [1]

$$ff'' + 2f''' = 0, \quad \text{and } \eta = 0, f = 0; f' = 0; \eta = \infty, f' = 1.  \tag{5}$$

The ordinary differential equation (5) are solved by the Runge-Kutta method for *f*. Then the velocity distribution of u and v are obtained by Eq.(4). Thus, the first and the second derivatives of u to y are

$$\frac{\partial u}{\partial y} = u_\infty f''(\eta)\sqrt{\frac{u_\infty}{vx}}, \quad \frac{\partial^2 u}{\partial y^2} = u_\infty (\frac{u_\infty}{vx}) f'''(\eta).  \tag{6}$$

The gradient of the total mechanical energy in the transverse direction is therefore

$$\frac{\partial E}{\partial y} = \rho u \frac{\partial u}{\partial y} = \rho u_\infty^2 \sqrt{\frac{u_\infty}{vx}} f'(\eta) f''(\eta) = \rho \frac{u_\infty^2}{x^2} (\frac{u_\infty x}{v})^{\frac{1}{2}} f'(\eta) f''(\eta), \qquad (7)$$

and the energy loss along the streamwise direction is

$$\frac{\partial H}{\partial x} = -\frac{\partial E}{\partial x} = -\mu \frac{\partial^2 u}{\partial y^2} = -\mu u_\infty (\frac{u_\infty}{vx}) f'''(\eta) = -\mu \frac{u_\infty}{x^2} (\frac{u_\infty x}{v}) f'''(\eta). \qquad (8)$$

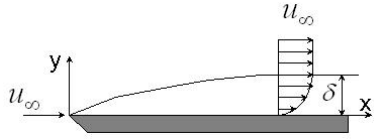

Fig. 1. The boundary layer along a flat plate at zero incidence.

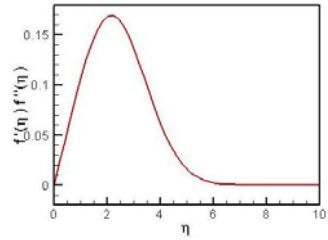

Fig. 2. Normalized energy gradient in transverse direction versus $\eta$.

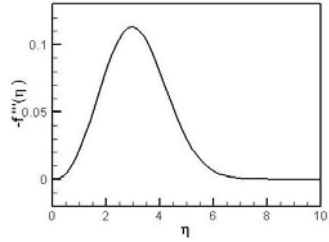

Fig. 3. Normalized energy loss along the streamwise direction versus $\eta$.

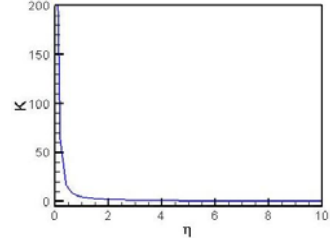

Fig. 4. Distribution of $K$ versus the transverse direction for boundary layer flow.

Introducing Eq.(7) and (8) into Eq.(1), the expression for K can be written as,

$$K = \frac{\partial E / \partial y}{\partial H / \partial x} = \text{Re}_x^{0.5} \frac{f' f''}{f'''}, \qquad (9)$$

where $\text{Re}_x = \rho u_\infty x / \mu$ is the local Reynolds number.

## 3. Results and Discussion

The variations of $\partial E / dy$ (normalized by $\rho(u_\infty^2 / x)(\text{Re}_x)^{0.5}$), $\partial H / \partial x$ (normalized by $\mu(u_\infty / x^2)\text{Re}_x$), and K (normalized by $\text{Re}_x$) versus the transverse coordinate $\eta$ are shown in Figs.2-4, respectively. It can be seen from Fig.2 that $\partial E / dy$ is zero at the wall and at the edge of the boundary layer, and it has a maximum within the boundary layer. These behaviours are the same as those for Poiseuille flows [4]. It also can be seen from Fig.3 that $\partial H/\partial x$ is zero at the wall and at the edge of the boundary layer and it has a maximum within the boundary layer, while $\partial H/\partial x$ is constant across the transverse direction for Poiseuille flows [4]. The behaviour of $\partial H/\partial x=0$ along the wall for boundary layer flow means that the fluid energy keeps to a constant along the wall. This fact is due to that the energy loss at the wall is exactly compensated by the external flow. It can be seen from Fig.4 that the magnitude of K approaches infinity towards the wall which is a singularity. This behaviour is mainly caused by the velocity profile which shows an inflexion at the wall, i.e., $\partial^2 u/\partial y^2 = 0$ due to $\partial p/\partial x = 0$. In terms of the stability criterion based on energy gradient, $K(v'_m/U) < Const$, the boundary layer flow is hence most dangerous at the wall for instability occurrence due to K being infinite. Since K is infinite at the wall, even a very small disturbance reaching the wall can make the flow become unstable and thus the transition to turbulence is generated for sufficiently large Re. This can be the mechanism of self-sustenance of wall turbulence in boundary layer flows [2, 6-8]. This is also the reason why turbulence is first generated at the wall as observed in the experiment [7-8]. When the level of external disturbance increases, the point of transition can be moved upstream according to the above criterion. The receptivity process is believed to act by affecting the value of $K(v'_m/U)$.

## 4. Conclusions

The energy gradient method is successfully employed to analyze the flow instability and turbulent transition in boundary layer flows. At sufficiently large $Re_x$, the stability criterion is first violated very near the wall at even very small perturbation level. This event serves as the origin of self-sustenance of wall turbulence. In comparison, in Poiseuille flows, a much larger disturbance is always needed to sustain the transition since the value of K is finite. The mechanism of receptivity to free-stream turbulence is possibly explained by the stability criterion.